\documentclass[12pt,twoside]{article}
\usepackage{fleqn,espcrc1}
\usepackage{graphicx}
\usepackage{epsfig}

\title{\vspace*{-1.0cm} \hfill {\rm MKPH-T-00-23}\\ \vspace{0.5cm}
The {\it E2/M1} and {\it C2/M1} ratios and form factors in
          $N\rightarrow \Delta$ transitions}

\author{L. Tiator, D. Drechsel, O. Hanstein\address{Institut f\"ur Kernphysik,
Universt\"at Mainz, D-55099 Mainz},
S.S. Kamalov\address{JINR Dubna, 141980 Moscow Region, Russia}
and
S.N. Yang\address{Department of Physics, National Taiwan University, Taipei 10617, Taiwan}}

\begin{document}

\maketitle

\begin{abstract}
A partial wave analysis of pion photoproduction has been obtained
in the framework of fixed-$t$ dispersion relations valid from
threshold up to 500 $MeV$. In the resonance region we have
precisely determined the electromagnetic properties of the
$\Delta(1232)$ resonance, in particular the E2/M1 ratio
$R_{EM}=(-2.5 \pm 0.1) \%$. For pion electroproduction recent
experimental data from Mainz, Bates and JLab for $Q^2$ up to 4.0
(GeV/c)$^2$ have been analyzed with two different models, an
isobar model (MAID) and a dynamical model. The E2/M1 ratios
extracted with these two models show, starting from a small and
negative value at the real photon point, a clear tendency to cross
zero, and become positive with increasing $Q^2$. This is a
possible indication of a very slow approach toward the pQCD
region. The C2/M1 ratio near the photon point is found as
$R_{SM}(0)=(-6.5 \pm 0.5) \%$. At high $Q^2$ the absolute value of
the ratio is strongly increasing, a further indication that pQCD
is not yet reached.

\end{abstract}

\section{INTRODUCTION}
\label{sec:intro}

The determination of the quadrupole excitation strength
$E_{1+}^{(3/2)}$ in the region of the $\Delta (1232)$ resonance has
been the aim of considerable experimental and theoretical activities.
Within the harmonic oscillator quark model, the $\Delta$ and the
nucleon are both members of the symmetrical 56-plet of $SU(6)$ with
orbital momentum $L = 0$, positive parity and a Gaussian wave
function in space. In this approximation the $\Delta$ may only be
excited by a magnetic dipole transition $M_{1+}^{(3/2)}$
\cite{Bec65}. However, in analogy with the atomic hyperfine
interaction or the forces between nucleons, also the interactions
between the quarks contain a tensor component due to the exchange of
gluons. This hyperfine interaction admixes higher states to the
nucleon and $\Delta$ wave functions, in particular $d$-state
components with $L = 2$, resulting in a small electric quadrupole
transition $E_{1+}^{(3/2)}$ between nucleon and $\Delta$
\cite{Kon80,Ger81,Isg82}. In addition quadrupole transitions are
possible by mesonic and gluonic exchange currents \cite{Buch97}.
Therefore an accurate measurement of $E_{1+}^{(3/2)}$ is of great
importance in testing the forces between the quarks and, quite
generally, models of nucleons and isobars.

The E2/M1 ratio, $R_{EM} =E_{1+}^{(3/2)}/M_{1+}^{(3/2)}$ has been
predicted to be in the range  $- 3\%\le R_{EM} < 0\%$ in the
framework of constituent quark \cite{Kon80,Isg82,Buch97,Dre84},
relativized quark \cite{Cap90,Cap92} and chiral bag models
\cite{Ber88}. Considerably larger values have been obtained in Skyrme
mo\-dels \cite{Wir87}. A first lattice QCD calculation resulted in a
small value with large error bars $(- 6 \% \le R_{EM} \le 12\%)$
\cite{Lei92}. However, the connection of the model calculations with
the experimental data is not evident. Clearly, the $\Delta$ resonance
is coupled to the pion-nucleon continuum and final-state interactions
will lead to strong background terms seen in the experimental data,
particularly in case of the small $E_{1+}$ amplitude. The question of
how to ''correct'' the experimental data to extract the properties of
the re\-so\-nance has been the topic of many theoretical
investigations. Unfortunately it turns out that the analysis of the
small $E_{1+}$ amplitude is quite sensitive to details of the models,
e.g. nonrelativistic vs. relativistic resonance denominators,
constant or energy-dependent widths and masses of the resonance,
sizes of the form factor included in the width etc. In other words,
by changing these definitions the meaning of resonance vs. background
changes, too.

In order to study the $\Delta$ deformation, pion photoproduction
on the proton has been measured by the LEGS collaboration
\cite{Bla97} at Brookhaven and by the A2 collaboration
\cite{Beck97} at Mainz using transversely polarized photons, i.e.
by measuring the polarized photon asymmetry $\Sigma$. In
particular, the cross section $d\sigma_{\parallel}$ for photon
polarization in the reaction plane turns out to be very sensitive
to the small $E_{1+}$ amplitude. Assuming for simplicity that only
the $P$-wave multipoles contribute, the differential cross section
is
\begin{equation}
\frac{d\sigma_{\parallel}}{d\Omega} = \frac{q}{k} (A_{\parallel} + B_{\parallel}
\cos \Theta_{\pi} + C_{\parallel} \cos^2 \Theta_{\pi}),
\end{equation}
where $q$ and $k$ are the pion and photon momenta and
$\Theta_{\pi}$ is the pion emission angle in the $c.m.$ frame.
Neglecting the (small) contributions of the Roper multipole
$M_{1-}$, one obtains \cite{Beck97}
\begin{equation}
C_{\parallel}/A_{\parallel} \approx 12 R_{EM} ,
\end{equation}
because the isospin $\frac{3}{2}$ amplitudes strongly dominate the cross
section for $\pi^{0}$ production.

In order to obtain the C2/M1 ratio and the form factors as
functions of $Q^2$ pion electroproduction has been studied. At
Mainz, Bonn, Bates and JLab different experiments have been
performed, without polarization as well as single and double
polarization.

While the experiments at Mainz and Bates measured at $Q^2\sim 0.1$
GeV$^2$ in order to get the C2/M1 ratio close to the photon point,
the JLab experiment was motivated by the possibility of
determining the range of momentum transfers where perturbative QCD
(pQCD) would become applicable. In the limit of $Q^2 \rightarrow
\infty$, pQCD predicts \cite{Brodsky81} that only
helicity-conserving amplitudes contribute, leading to
 $R_{EM} = E_{1+}^{(3/2)}/M_{1+}^{(3/2)} \rightarrow 1$ and
 $R_{SM} = S_{1+}^{(3/2)}/M_{1+}^{(3/2)} \rightarrow const$.

\section{PION PHOTOPRODUCTION}
\label{sec:real}

Starting from fixed-$t$ dispersion relations for the invariant
amplitudes of pion photoproduction, the projection of the multipole
amplitudes leads to a well known system of integral equations,
\begin{equation} \label{inteq}
\mbox{Re}{\cal M}_{l}(W) = {\cal M}_{l}^{\mbox{\scriptsize P}}(W)
+ \frac{1}{\pi}\sum_{l'}{\cal P}\int_{W_{\mbox{\scriptsize thr}}}^{\infty}
K_{ll'}(W,W')\mbox{Im}{\cal M}_{l'}(W')dW',
\end{equation}
where ${\cal M}_l$ stands for any of the multipoles
$E_{l\pm}, M_{l\pm},$ and ${\cal M}_{l}^{\mbox{\scriptsize P}}$ for the
corresponding (nucleon) pole
term. The kernels $K_{ll'}$ are known, and the real and imaginary parts of the
amplitudes are related by unitarity. In the energy region below two-pion
threshold, unitarity is expressed by the final state theorem of
Watson,
\begin{equation} \label{watson}
{\cal M}_l^I (W) = \mid {\cal M}_{l}^{I} (W)\mid
e^{i(\delta_{l}^{I} (W) + n\pi)},
\end{equation}
where $\delta_{l}^{I}$ is the corresponding $\pi N$ phase shift
and $n$ an integer. We have essentially followed the method of
Schwela et al \cite{Sch69,Pfe72} to solve Eq. (\ref{inteq}) with
the constraint (\ref{watson}). In addition we have taken into
account the coupling to some higher states neglected in that
earlier reference. At the energies above two-pion threshold up to
$W = 2$ GeV, Eq.~(\ref{watson}) has been replaced by an Ansatz
based on unitarity \cite{Sch69}. Finally, the contribution of the
dispersive integrals from $2$ GeV to infinity has been replaced by
$t$-channel exchange, parametrized by certain fractions of $\rho$-
and $\omega$-exchange. Furthermore, we have to allow for the
addition of solutions of the homogeneous equations to the coupled
system of Eq.~(\ref{inteq}). The whole procedure introduces 9 free
parameters, which have been determined by a fit to the data.
\cite{Han98}

In Figure 1 we show the $P_{33}$ multipoles $M^{3/2}_{1+}$ and
$E^{3/2}_{1+}$. Our dispersion theoretical analysis (solid line)
agrees very well with our single-energy fit and with the
single-energy fit of Beck et al. \cite{Beck00}. The only
systematic deviation becomes visible in the electric multipole
above the resonance position. This can be due either to our
truncation of partial waves or to systematics in the experiment at
the highest energies. In a new experiment a full angular coverage
of the differential cross section for $p(\gamma,\pi^0)p$ over a
wide range of energy from threshold up to $E^{lab}_\gamma =440
MeV$ has been taken and in the near future a new and very precise
multipole analysis should also clarify this small deviation.
\begin{figure}[htbp]
\centerline{\epsfig{file=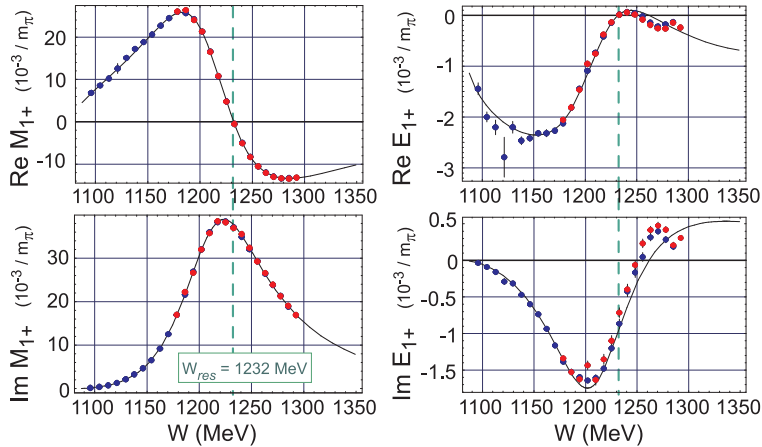,width=10cm}}
\caption[ ]{ $P_{33}$ amplitudes
  $M_{1+}^{3/2}$ and $E_{1+}^{3/2}$ for pion photoproduction in the
  $\Delta(1232)$ region. The black points are our single-energy fits,
  the grey points are taken from Beck et al. \cite{Beck00}. In most cases
  they exactly overlap.}
\label{p33}
\end{figure}
According to the Watson theorem, at least up to the two-pion
threshold, the ratio $E_{1+}^{(3/2)}/M_{1+}^{(3/2)}$ is a real
quantity. However, it is not a constant but even a rather strongly
energy dependent function. If we determine the resonance position
as the point, where the phase
$\delta_{1+}^{(3/2)}(W=M_\Delta)=90^\circ$, we can define the
so-called "full" ratio
\begin{equation}
R_{EM} = \left. \frac{E_{1+}^{(3/2)}}{M_{1+}^{(3/2)}}\right|_{W=M_\Delta}
       = \left. \frac{\mbox{Im}E_{1+}^{(3/2)}}{\mbox{Im}M_{1+}^{(3/2)}}
       \right|_{W=M_\Delta}\,.
\end{equation}
We note that this ratio is identical to the ratio obtained with
the $K$-matrix at the $K$-matrix pole $W=M_\Delta$. This can be
seen by using the relation between the $T$- and the $K$-matrix,
\begin{equation}
T=K\cos\delta e^{i\delta} \quad \mbox{and consequently} \quad
K=\mbox{Re}T+\mbox{Im}T\, tan\delta\,.
\end{equation}
Therefore, at $W=M_\Delta$ we find
\begin{equation}
K(E_{1+}^{(3/2)})/K(M_{1+}^{(3/2)})=
\mbox{Im}E_{1+}^{(3/2)}/\mbox{Im}M_{1+}^{(3/2)}=R_{EM}\,.
\end{equation}
The recent, nearly model-independent value of the Mainz group at
$W =M_{\Delta}=1232$ MeV is $(-2.5 \pm 0.1 \pm 0.2)\%$
\cite{Beck00} is in excellent agreement with our dispersion
theoretical calculation that gives $(-2.54\pm 0.10)\%$, see Table
1.
\begin{table}[htbp]
\caption{\small E2/M1 ratios for $Q^2$=0 from different analyses.}
\begin{tabular}{l|l}
\hline
$R_{EM} [\%]$ & Reference \\ \hline
$ -2.54 \pm 0.10 $ & Hanstein et al. \cite{Han98} \\
$ -2.5 \pm 0.1_{stat.} \pm 0.2_{syst.} $ & Beck et al. \cite{Beck00} \\
$ -3.0 \pm 0.3_{stat.+syst.} \pm 0.2_{mod.} $ & Blanpied et al. \cite{Bla97}\\
$ -1.5 \pm 0.5 $ & Arndt et al. \cite{Arn97} \\
$ -3.19 \pm 0.24 $ & Davidson et al. \cite{Dav97} \\ \hline
$ -2.5 \pm 0.5 $ & PDG 2000 estimate \cite{PDG00} \\ \hline
\end{tabular}
\end{table}
As it was demonstrated in different approaches \cite{Han98,Dav99},
the precise E2/M1 ratio is very sensitive to the specific database
used in the fit. Therefore, the SAID value, obtained with the full
database is rather low ($-1.5 \%$) and the values obtained with
the LEGS differential cross sections are twice as large, around
$-3\%$.

The ratio so far discussed above, is the ratio at the K-matrix
pole on the real energy axis. In scattering theory, the T-matrix
pole in the complex plane, however, is more fundamental. The
analytic continuation of a resonant partial wave as function of
energy into the second Riemann sheet should generally lead to a
pole in the lower half-plane. A pronounced narrow peak reflects a
time-delay in the scattering process due to the existence of an
unstable excited state. This time-delay is related to the speed
$SP$ of the scattering amplitude $T$, defined by \cite{Hoe92}
\begin{equation}
SP(W) = \left\vert \frac{dT(W)}{dW}\right\vert ,
\end{equation}
where $W$ is the total $c.m.$ energy. In the vicinity of the resonance
pole, the energy dependence of the full amplitude $T = T_{B} + T_{R}$
is determined by the resonance contribution,
\begin{equation}\label{T_res}
  T_{R} (W) = \frac{r\Gamma_{R} e^{i\phi}}{M_{R}- W-
    i\Gamma_{R}/2}\,\,,
\end{equation}
while the background contribution $T_B$ should be a smooth
function of energy, ideally a constant. We note in particular that
$W_R = M_{R}- i\Gamma_{R}/2$ indicates the position of the
resonance pole in the complex plane, i.e. $M_{R}$ and $\Gamma_{R}$
are constants and differ from the energy-dependent widths, and
possibly masses, derived from fitting certain resonance shapes to
the data. In the limit where the derivative of the smooth
background can be neglected, the speed takes the simple form
\begin{equation}\label{Speed}
  SP(W) = \frac{r\Gamma_{R}}{(M_{R}- W)^2+
    \Gamma_{R}^2/4}\,.
\end{equation}
From this form, the position of the pole as well as the absolute
magnitude of the residue can be easily obtained. Furthermore, in
our dispersion approach we have also checked the validity of the
assumption to neglect the background and found that this procedure
works very well for the Delta resonance.

Applying this technique to our $P_{33}$ amplitudes we find the pole at
$W_{R} = M_{R} - i \Gamma_{R}/2 = (1211 - 50 i)$ MeV in excellent
agreement with the results obtained from $\pi N$ scattering, $M_{R}=
(1210\pm 1)$ MeV and $\Gamma_{R} = 100$ MeV \cite{Hoe92}.  The complex
residues and the phases are obtained as $r_E=1.23\cdot
10^{-3}/m_{\pi}, \phi_E=-154.7^{\circ}, r_M=21.16\cdot
10^{-3}/m_{\pi}$ and $\phi_M=-27.5^{\circ}$, yielding a complex ratio
of the residues
\begin{equation}
R_{\Delta} = \frac{r_{E} e^{i\phi_{E}}}{r_{M} e^{i\phi_{M}}}
= - 0.035 - 0.046 i.
\end{equation}
While the experimentally observed ratio $R_{EM}$ is real and very
sensitive to small changes in energy, the ratio $R_{\Delta}$ is a
complex number defined by the residues at the pole, therefore, it does
not depend on energy.

It should be noted, however, that a resonance without the
accompanying background terms is unphysical, in the sense that only
the sum of the two obeys unitarity. Furthermore we want to point out
that the speed-plot technique does not give information about the
strength parameters of a "bare" resonance, i.e. in the case where the
coupling to the continuum is turned off. Both the pole position and
the residues at the pole will change for such a hypothetical case, but
the exact values for the "bare" resonance can only be determined by a
model calculation and as such will depend on the ingredients of the
model.

\section{PION ELECTROPRODUCTION}
\label{sec:virtual}

 In the dynamical approach to pion photo- and electroproduction
\cite{Yang85}, the t-matrix is expressed as
\begin{eqnarray}
t_{\gamma\pi}(E)=v_{\gamma\pi}+v_{\gamma\pi}\,g_0(E)\,t_{\pi
N}(E)\,, \label{eq:tgamapi}
\end{eqnarray}
where $v_{\gamma\pi}$ is the transition potential operator for
$\gamma^*N \rightarrow \pi N$, and $t_{\pi N}$ and $g_0$ denote
the $\pi N$ t-matrix and free propagator, respectively, with $E
\equiv W$ the total energy in the CM frame. A multipole
decomposition of Eq. (\ref{eq:tgamapi}) gives the physical
amplitude in channel $\alpha$~\cite{Yang85},
\begin{eqnarray}
t_{\gamma\pi}^{(\alpha)}(q_E,k;E+i\epsilon)
&=&\exp{(i\delta^{(\alpha)})}\,\cos{\delta^{(\alpha)}}
\nonumber\\&\times&
\left[v_{\gamma\pi}^{(\alpha)}(q_E,k) + P\int_0^{\infty}
dq' \frac{q'^2R_{\pi
N}^{(\alpha)}(q_E,q';E)\,v_{\gamma\pi}^{(\alpha)}(q',k)}{E-E_{\pi
N}(q')}\right], \label{eq:backgr}
\end{eqnarray}
where $\delta^{(\alpha)}$ and $R_{\pi N}^{(\alpha)}$ are the $\pi
N$ scattering phase shift and reaction matrix in channel $\alpha$,
respectively; $q_E$ is the pion on-shell momentum and $k=|{\bf
k}|$ is the photon momentum. The multipole amplitude in Eq.
(\ref{eq:backgr}) manifestly satisfies the Watson theorem and
shows that the $\gamma\pi$ multipoles depend on the half-off-shell
behavior of the $\pi N$ interaction.

In a resonant channel like (3,3) in which the $\Delta(1232)$ plays
a dominant role, the transition potential $v_{\gamma\pi}$ consists
of two terms,
\begin{eqnarray}
v_{\gamma\pi}(E)=v_{\gamma\pi}^B + v_{\gamma\pi}^{\Delta}(E)\,,
\label{eq:tranpot}
\end{eqnarray}
where $v_{\gamma\pi}^B$ is the background transition potential and
$v_{\gamma\pi}^{\Delta}(E)$ corresponds to the contribution of the
bare $\Delta$.

It is well known that for a correct description of the resonance
contributions we need, first of all, a reliable description of the
nonresonant part of the amplitude.

In the new version of MAID (MAID2000), the $S$, $P$, $D$ and $F$
waves of the background contributions are complex numbers defined in
accordance with the K-matrix approximation,
\begin{equation}
 t_{\gamma\pi}^{B,\alpha}({\rm MAID})=
 \exp{(i\delta^{(\alpha)})}\,\cos{\delta^{(\alpha)}}
 v_{\gamma\pi}^{B,\alpha}(W,Q^2).
\label{eq:bg00}
\end{equation}
From Eqs. (\ref{eq:backgr}) and  (\ref{eq:bg00}), one finds that
the difference between the background terms of MAID and of the
dynamical model is that off-shell rescattering contributions
(principal value integral) are not included in MAID. To take
account of the inelastic effects at the higher energies, we
replace $\exp{(i\delta^{(\alpha)})} \cos{\delta^{(\alpha)}} =
\frac 12 (\exp{(2i\delta^{(\alpha)})} +1)$ in Eqs.
(\ref{eq:backgr}) and (\ref{eq:bg00}) by $\frac 12
(\eta_{\alpha}\exp{(2i\delta^{(\alpha)})} +1)$, where
$\eta_{\alpha}$ is the inelasticity. In our actual calculations,
both the $\pi N$ phase shifts $\delta^{(\alpha)}$ and inelasticity
parameters $\eta_{\alpha}$ are taken from the analysis of the GWU
group \cite{Arn97}.

Following  Ref.~\cite{Maid},  we assume a Breit-Wigner form for the
resonance contribution ${\cal A}^{R}_{\alpha}(W,Q^2)$ to the total
multipole amplitude,
\begin{equation}
{\cal A}_{\alpha}^R(W,Q^2)\,=\,{\bar{\cal A}}_{\alpha}^R(Q^2)\,
\frac{f_{\gamma R}(W)\Gamma_R\,M_R\,f_{\pi
R}(W)}{M_R^2-W^2-iM_R\Gamma_R} \,e^{i\phi}, \label{eq:BW}
\end{equation}
where $f_{\pi R}$ is the usual Breit-Wigner factor describing the
decay of a resonance $R$ with total width $\Gamma_{R}(W)$ and
physical mass $M_R$. The expressions for $f_{\gamma R}, \, f_{\pi R}$
and $\Gamma_R$ are given in Ref.~\cite{Maid}. The phase $\phi(W)$ in
Eq. (\ref{eq:BW}) is introduced to adjust the phase of the total
multipole to  equal  the corresponding $\pi N$  phase shift
$\delta^{(\alpha)}$. Because  $\phi=0$ at resonance, $W=M_R$, this
phase does not affect the $Q^2$ dependence of the $\gamma N R$
vertex.

In the dynamical model of Ref. \cite{KY99}, a scaling assumption
was made concerning the (bare) form factors ${\bar{\cal
A}}_{\alpha}^\Delta(Q^2)$, namely, that all of them have the same
$Q^2$ dependence,
\begin{eqnarray}
{\bar{\cal A}}_{\alpha}^{\Delta}(Q^2)={\bar{\cal
A}}_{\alpha}^{\Delta}(0) \frac{ k}{k_W}\,F(Q^2),\,
\end{eqnarray}
where $\alpha = M, E,$ and $S$, $k_W = (W^2 - m_N^2)/2W$, ${
k}^2=Q^2+((W^2-m_N^2-Q^2)/2W)^2$, and $F$ is normalized to $F(0) =
1$. The values of ${\bar{\cal A}}_M^{\Delta}(0)$ and ${\bar{\cal
A}}_E^{\Delta}(0)$ were determined by fitting to the multipoles
obtained in the recent analyses of the Mainz \cite{Han98} and GWU
\cite{VPI97} groups. The $Q^2$ evolution of the form factor $F$
was assumed to take the form $ F(Q^2)=(1+\beta\,Q^2)\,e^{-\gamma
Q^2}\,G_D(Q^2),$ where $G_D(Q^2)=1/(1+Q^2/0.71)^2$ is the usual
dipole form factor. The parameters $\beta$ and $\gamma$ were
determined by fitting ${\bar{\cal A}}_{M}^{\Delta}(Q^2)$ to the
data for $G_M^*$ defined by \cite{Maid,KY99},
\begin{eqnarray}
M_{1+}^{(3/2)}(M_{\Delta},Q^2)=\frac {k}{m_N}
\sqrt{\frac{3\alpha_{em}}{8\Gamma_{exp} q_{\Delta}}}\, G_M^*(Q^2),
\label{eq:gmform2}
\end{eqnarray}
where $\alpha_{em}=1/137$, $\Gamma_{exp}=115$ MeV, and
$q_{\Delta}$ is the pion momentum at $W=M_\Delta$. With the
relation ${\bar{\cal A}}_E^{\Delta}(0)= {\bar{\cal
A}}_S^{\Delta}(0)$, the ratios $R_{EM}$ and $R_{SM}$ between the
full multipoles were then evaluated \cite{KY99} and found to agree
with the values extracted in Ref. \cite{Frolov99}.

In the present analysis, we do not impose the scaling assumption and
write, for electric ($\alpha=E$) and
Coulomb ($\alpha=S$) multipoles,
\begin{eqnarray}
{\bar{\cal A}}_{\alpha}^{\Delta}(Q^2)=X_{\alpha}^{\Delta}(Q^2)\,{\bar{\cal
A}}_{\alpha}^{\Delta}(0) \frac{ k}{k_W}\,F(Q^2),
\end{eqnarray}
with $X_{\alpha}^{\Delta}(0) = 1$,  and we allow  both $X_E$ and
$X_S$ to be determined by the experiment.

\begin{figure}[bht]
\centerline{\epsfig{file=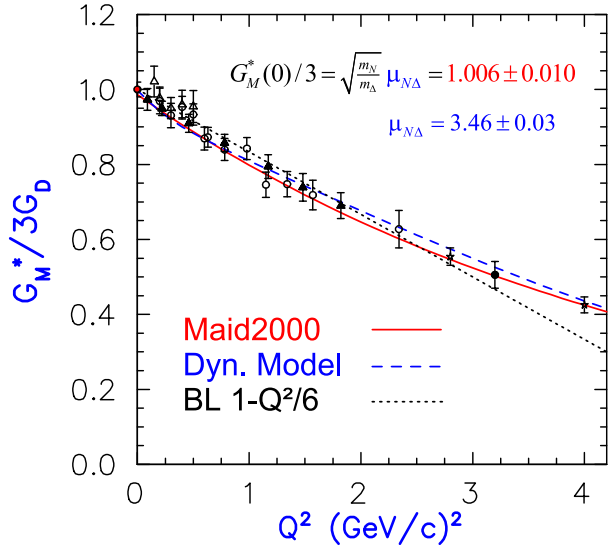,width=8cm}} \vspace{-0.5cm}
\caption{ The $Q^2$ dependence of the magnetic $N \rightarrow
\Delta$ transition form factor $G_M^*$ divided by three times the
nucleon dipole form factor. The solid and dashed curves are the
results of the MAID and dynamical model analyses, respectively.
The dotted line shows the simple fit of Ref. \cite{Lag88}. The
data at $Q^2$=2.8 and 4.0 $(GeV/c)^2$ are from Ref.
\cite{Frolov99}. For other data see Ref. \cite{Kamalov00}. }
\end{figure}
\label{fig:gm} The dynamical model and MAID are used to analyze
the recent JLab differential cross section data on $p(e,e'p)\pi^0$
at high $Q^2$. All measured data, 751 points at $Q^2$=2.8 and 867
points at $Q^2$=4.0 (GeV/c)$^2$ covering the entire energy range
$1.1 < W < 1.4$ GeV, are included in our global fitting procedure
using the MINUIT code and we obtain a very good fit to the
measured differential cross sections. Our results for the $G_M^*$
form factor are shown in Figure 2. Here the best fit is obtained
with $\gamma=0.21$ (GeV/c)$^{-2}$ and $\beta=0$ in the case of
MAID, and $\gamma=0.40$ (GeV/c)$^{-2}$ and
$\beta=0.52$(GeV/c)$^{-2}$ in the case of the dynamical model. It
is worth noting that in the definition of Eq.~(\ref{eq:gmform2}),
$G_M^*(0)/3$ takes a value of 1 to an accuracy of $1\%$. This very
precise value is extracted from the recent Mainz experiment
\cite{Beck00}. With this number we can also determine a very
precise $N \rightarrow \Delta$ magnetic transition moment,
$\mu_{N\Delta}=3.46 \pm 0.03$ in units of nuclear magnetons.

\begin{figure}[bht]
\centerline{\epsfig{file=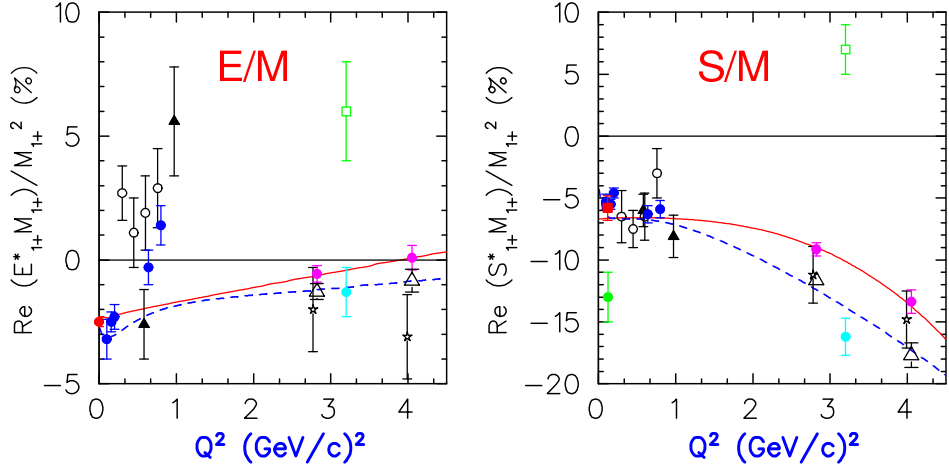,height=12.5cm,angle=-90}}
\caption{ The $Q^2$ dependence of the ratios $R_{EM}^{(p\pi^0)}$
and $R_{SM}^{(p\pi^0)}$ at $W=1232$ MeV. The solid and dashed
curves are the MAID and dynamical model results, respectively.
Experimental data at $Q^2=0$ from Ref. \cite{Beck97}, data at
$Q^2$=0.1, 0.16, 0.2, 0.64, 0.8 $GeV^2$ from Ref. \cite{Gothe00}
and data at $Q^2$=2.8 and 4.0 $GeV^2$ from Ref. \cite{Frolov99}
(stars). Our results at $Q^2$=2.8 and 4.0 $GeV^2$, including their
error bars, are obtained from MAID ($\bullet$) and the dynamical
model ($\bigtriangleup$) analyses. The points at $Q^2=3.2$ GeV$^2$
are obtained from DESY data \cite{Haidan79} by Ref.
\cite{Burkert95} (squares) and MAID ($\bullet$). For other data
see Ref. \cite{Kamalov00}.} \label{fig:ratios}
\end{figure}
\begin{table}[htbp]
\caption{\small Our results for the magnetic transition form
factor $G_M^*$ and for the ratios $R_{EM}$ and $R_{SM}$, at
$Q^2$=2.8 (upper row) and 4.0 (lower row) (GeV/c)$^2$, extracted
from a global fit to the data with MAID and the dynamical model as
discussed in the text. Results from Ref. \protect\cite{Frolov99}
are listed for comparison. Ratios are given in (\%).}
\begin{tabular}{c|c|ccc}
\hline
 ratios & $Q^2(GeV^2)$    &  MAID  &  DM  & Ref. \cite{Frolov99}
 \\\hline
$ G_M^*\times 100$
                 & 2.8 & $  6.78\pm 0.05 $ & $ 7.00\pm 0.04$
& $ 6.9\pm 0.4$ \\
                 & 4.0 &  $  2.86\pm 0.02 $ & $ 3.04\pm 0.02$
& $ 2.9\pm 0.2$\\ \hline
$R_{EM}$  & 2.8 &$ -0.56\pm 0.33 $ & $-1.28\pm 0.32$ &
$-2.00\pm 1.7$\\ & 4.0 &$  0.09\pm 0.50 $ & $-0.84\pm 0.46$ &
$-3.1\pm 1.7$ \\
\hline $R_{SM}$  & 2.8 &$ -9.14\pm 0.54 $ & $-11.65\pm 0.52$ &
$-11.2\pm 2.3$\\
                 & 4.0 & $-13.37\pm 0.95 $ & $-17.70\pm 1.0 $ &
$-14.8\pm 2.3 $ \\ \hline
\end{tabular}
\end{table}
We have also re-analyzed older DESY \cite{Haidan79} data measured
at $Q^2=3.2 (GeV/c)^2$ and found significantly different results
for the ratios compared to a previous analysis \cite{Burkert95}.
However, since these data also give a $G_M^*$ value $20 \%$ below
our fit curves in Figure 2, we did not include them in our fits of
the ratios.

In a similar way we also analyzed recent Bates measurements
\cite{Mertz99} for unpolarized differential cross sections,
$R_{LT}$ response function and $A_{LT}$ asymmetry for
$p(e,e'\pi^0)p$ at $Q^2=0.126 (GeV/c)^2$ and obtained the
following preliminary values for the E2/M1 and C2/M1 ratios
\begin{equation}
R_{EM}=(-2.1\pm 0.2)\% \quad \mbox{and} \quad R_{SM}=(-6.3\pm 0.2)\% \,.
\end{equation}
Finally, in a double polarization experiment at Mainz
\cite{Schmieden99},  measuring the recoil polarization of the
proton, $p(\vec{e},e'\vec{p})\pi^0$ at $Q^2=0.121 (GeV/c)^2$, a
preliminary value of
\begin{equation}
R_{SM}=(-5.8\pm 1.0)\% \,.
\end{equation}
could be extracted in a rather model-independent way from the
x-component of the recoil polarization $P_x$, which is very
sensitive to the resonant $S_{1+}$ multipole.

Our extracted values for $R_{EM}$ and $R_{SM}$ and a comparison
with the results of Ref.~\cite{Frolov99} are presented in Table 2
and shown in Figure 3. The main difference between our results and
those of Ref.~\cite{Frolov99} is that our values of $R_{EM}$ show
a clear tendency to cross zero and change sign as $Q^2$ increases.
This is in contrast with the results obtained in the original
analysis \cite{Frolov99} of the data which concluded that $R_{EM}$
would stay negative and tend toward more negative values with
increasing $Q^2$.

\section{CONCLUSIONS}
\label{sec:concl}

In the framework of fixed-t dispersion relations with the new and
very precise data obtained at MAMI in Mainz we have obtained a new
partial wave analysis for pion photoproduction. The uncertainties
in most multipoles could be considerably improved compared to
previous analyses. At the resonance position, where the phase
passes $90^{\circ}$, we obtain an E2/M1 ratio of $R_{EM}=(-2.5 \pm
0.1) \%$. At the pole in the complex plane we obtain the ratio of
the resonant electric and magnetic multipoles as $R_{\Delta} = -
0.035 - 0.046 i$.  This is a further model-independent ratio that
can be determined in any analysis or calculation of pion
photoproduction.

For pion electroproduction, we have re-analyzed the recent JLab
data for electroproduction of the $\Delta(1232)$ resonance via
$p(e,e'p)\pi^0$ with two models for pion electroproduction, both
of which give excellent descriptions of the existing data. In
contrast to previous findings, our models indicate that $R_{EM}$,
starting from a small and negative value at the real photon point,
actually exhibits a clear tendency to cross zero and change sign
as $Q^2$ increases. It will be most interesting to have data at
yet higher momentum transfer in order to see whether such a trend
continues, which would be a sign for a rather slow approach
towards the pQCD region. Furthermore, the absolute value of
$R_{SM}$ is strongly increasing, which indicates that the pQCD
prediction of $R_{SM}\rightarrow constant$ is not yet reached.

\section{ACKNOWLEDGMENTS}
We wish to thank R. Beck, R. Leukel, H. Schmieden, R. Gothe and C.
Papanicolas for their contribution on the experimental data. This
work was supported in part by NSC under Grant No.
NSC89-2112-M002-038, by Deutsche Forschungsgemeinschaft (SFB443)
and by a joint project NSC/DFG TAI-113/10/0.

\end{document}